\documentclass[english,twoside,a4paper,10pt]{article}

\usepackage[latin1]{inputenc}
\usepackage[T1]{fontenc}
\usepackage{amsmath}
\usepackage{amsfonts}
\usepackage{graphicx}
\usepackage{subfigure}
\usepackage{a4wide}
\usepackage{amssymb}
\usepackage{fancyhdr}
\usepackage{mathrsfs}
\usepackage[toc,page]{appendix}

\begin{document}

\title{Action growth for  black holes in modified  gravity}

\author{ 
Lorenzo Sebastiani\footnote{E-mail address: lorenzo.sebastiani@unitn.it
},\,\,\,
Luciano Vanzo\footnote{E-mail address: luciano.vanzo@unitn.it},\,\,\,
Sergio Zerbini\footnote{E-mail address: zerbini@science.unitn.it}\\
\\
\begin{small}
Dipartimento di Fisica, Universit\`a di Trento,Via Sommarive 14, 38123 Povo (TN), Italy
\end{small}\\
\begin{small}
TIFPA - INFN,  Via Sommarive 14, 38123 Povo (TN), Italy
\end{small}
}

\date{}

\maketitle

\abstract{The general form of the action growth for a large class of static black hole  solutions in 
modified gravity which includes $F(R)$-gravity models is computed. The cases of black hole solutions with non constant Ricci
scalar are also considered, generalizing the results previously found and valid only for  black holes with constant Ricci scalar.
An argument is put forward to provide a physical interpretation of the results, which seem tightly connected with the generalized second law of black hole thermodynamics.}

\section{Introduction}

Recently, Brown {\it et al.} proposed an interesting conjecture in the AdS/CFT framework,
according to which the quantum computational complexity of a holographic  state may be inferred from the classical action related to  a specific region in the bulk~\cite{Sus}. Such a proposal has been checked in the context of the Anti de Sitter (AdS) black holes (BHs) in General Relativity (GR),  and this is an  interesting test  for the CA (complexity/action) duality~\cite{Susskind}. This conjecture is a refined version of a previous one which states that the complexity is dual to the spatial volume of a maximal slice behind the horizon~\cite{Sus0}. Since the properties of the black hole interior are represented on the holographic boundary, it is possible to find 
the boundary state by computing the classical action of the space-time region inside the BH (in the so called ``Wheeler-DeWitt patch'', see Ref.~\cite{Myers} for a detailed geometrical 
analysis of the issue). After calculating the growth of the complexity at the late time, it is found that in the case of neutral black holes the action growth is bounded by a term proportional to the BH energy. 

In modified theories of gravity several attempts have been made in order 
to calculate the action growth for neutral and charged AdS black holes, see for example Refs.~\cite{Caibound, FRbound,Wang}.

In this paper, our aim is to investigate the action growth in  the case of the  black holes within a class of modified gravity. We will be mainly interested in  $F(R)$-theories of gravity, where the action is given by a general function of the Ricci scalar $R$. Such models  represent the simplest
generalization of the Einstein's theory, and, in general, they  admit the existence of Schwarzschild dS/AdS black holes, namely solutions with constant Ricci curvature. Beside these ``trivial'' black 
hole solutions, we will present the computation of the action growth associated with non trivial  black hole (vacuum) solutions with non constant Ricci curvature,
found in  Refs.~\cite{CB, Mutamaki, Capozziello, iran, Seba, cogno}. Some of these  static solutions represent ``dirty BHs''~\cite{Visser}, namely ones in which the (00)- and (11)-metric components are related as $g_{00}g_{11} \neq -1 $. They typically involve scalar hairs.

The thermodynamical interpretation for such  BHs solutions is still an open issue (see for instance Refs.~\cite{Deserenergy, Abreu, Cai, su, CH})
and the relation between the action growth and the BH energy in $F(R)$-gravity should be careful considered.
For our purposes, we will make use of the fact that in most cases and within $F(R)$-gravity, the BH energy may be
obtained by deriving the First Law of BH thermodynamics from the equations of motion~\cite{Gorbu}. In fact, when only one integration constant appears in the  solution, it is possible to identify it with the Killing energy of the black hole itself.

For the black holes with constant Ricci curvature, we confirm the results previously obtained. For dirty black holes with non constant Ricci scalar,  the so called Kodama-Hayward energy
appears in the action growth. Finally, we also  investigate the action growth for a modified gravity model with an additional term based on the Weyl tensor, which is not belonging to the  $F(R)$-class.

Quite apart from  computations, it will also be important to assess 
the validity of the  conditions allowing us to restrict attention to spherically
symmetric solutions beyond the obvious demand of simplicity and the advantage of working with exact solutions, and to relate
the action grow with the physics of black hole evaporation. In this context, the more important property is the grows being proportional
to the internal energy of the black hole. We will show that this is equivalent, for neutral non rotating black holes, to the simultaneous validity of the generalized second
law together with the  Pendry's inequality \cite{pendry-1983} characterizing the information rate of a single communication channel, 
whose exact definition in general depends on the physical character of the information carriers and the medium by which they 
propagate\footnote{We are using these terms in the sense of Shannon's communication theory\cite{cover}. One may conveniently think of 
a one-dimensional channel as an optical fiber.}. We recall that the generalized second law stipulates that the entropy of the black
hole plus the one carried away by the Hawking radiation should satisfy the inequality
\begin{equation}
0\leq \dot{S}_{BH}+\dot{S}_{rad}\,,
\end{equation}
where the dot denotes the time derivative\footnote{Here time derivatives are taken with respect to retarded 
coordinate time, or equivalently to time at infinity.},
while what Pendry says (adapted in a form suitable to us) is that for a channel fed by power $P$, we have 
\begin{equation}
\dot{S}_{+}\leq \left(\frac{\pi P}{3}\right)^{1/2}\,,\label{PI2}
\end{equation}
where $\dot{S}_{+}$ is the entropy flow along the channel. Identifying $S_{+}$ with $S_{rad}$ and $P=-\dot{E}_{BH}$, both are satisfied by the black holes and together would imply that the action grow scales with the internal energy, so we may say that the neutral black hole is a kind of one-dimensional information channel in the sense specified by Pendry, as was shown long ago by Bekenstein  by other means\cite{beken2001}. Adopting the CA conjecture, one may conclude that the rate of complexity grow of (the boundary horizon state) of the black hole cannot be more than twice its thermodynamical energy.

The rest of the paper is organized in the following way. In Section {\bf 2}, the equations of motion for static spherical symmetric (SSS) metric of  $F(R)$-gravity are calculated, starting from a 
suitable action in which  the associated boundary term has been taken into account. 
In  Section {\bf 3} we present a derivation of the First law of BH thermodynamics which allows to obtain the BH Killing energy in the
framework of $F(R)$-gravity. In Section {\bf 4} the general formalism for the evaluation of the action growth in $F(R)$-gravity is presented and applied to the black holes previously 
introduced. We use a simpler approach making full use of the assumed spherical symmetry. For a full treatment in general relativity in anti-de Sitter space, see the
recent comprehensive paper of D.~Carmi {\it et al.}~\cite{Carmi:2017jqz}. Section {\bf 5} is devoted to the calculation of the action growth for a BH solution in a Weyl model of modified gravity. After these rather technical sections, in Section {\bf 6} we give a physical discussion of the results thereby obtained. The conclusions and final remarks are given in Section {\bf 7}, while in the Appendixes 
 the explicit calculations of the boundary terms of the action in $F(R)$- and Weyl-gravity are presented.
 
In this work we use units of $k_{\mathrm{B}} = c = \hbar = 1$.

\section{Action and equations of motion in $F(R)$-gravity}

To begin with, we recall that the action for a generic modified gravity model depending only on the scalar Ricci curvature in the vacuum and in four dimensions may be written as (see for example ~\cite{Od,va,de} ), 
\begin{equation}
 I=\int_\mathcal{M} d^4 x\sqrt{-g}F(R)\,, \label{action0}
\end{equation}
where $\mathcal M$ is a four-dimentional space-time manifold with boundary $\partial \mathcal M$, $g$ is
the determinant of the metric tensor $g_{\mu\nu}(x^\mu)$, and $F(R)$ is a function of the Ricci scalar $R$.

As in GR, in order to deal with a proper well posed
variational problem for the metric tensor~\cite{Gibbons}, one needs to subtract to the Lagrangian a suitable boundary  term. 
In the so called Jordan frame (JF),  one has to work with the following action~\cite{Buch,Ma,Fa,Franca, guarnizo, deruelle00, deruelle0, deruelle},
\begin{equation}
\hat{I}=\int_\mathcal{M} d^4 x\sqrt{-g}F(R)-2 \int_{\partial \mathcal M} d^3 x\sqrt{-h}F'(R)K \,,\label{genBT}
\end{equation}
where $K$ is the trace of the extrinsic curvature related to $\partial M$ and $h$ is the trace of the three-dimensional induced metric $h_{ij}(x^i)$. Usually the boundary has topology $S^2\times R$ and is foliated by two-spheres. The signature can be either time-like or null, but not space-like. In the null case there are some unresolved ambiguities\cite{Myers}. If it is not orthogonal to the space-time foliation in the Hamiltonian formulation then suitable bolt terms have to be added, along the lines discussed in \cite{Hawking:1996ww} in GR, for example.
The field equations can be derived and one gets
\begin{equation}
F'(R)R_{\mu \nu}-\frac{1}{2}F(R)g_{\mu\nu}-\left(\nabla_\mu \nabla_\nu-g_{\mu\nu}\nabla_\alpha \nabla^\alpha\right)F'(R)=0\,,\label{keyeq}
\end{equation}
where $\nabla_\mu$ is the covariant derivative associated to the metric tensor $g_{\mu\nu}(x^\mu)$ and the prime denotes the derivative with respect to the Ricci scalar.
As well known, these set of above differential equations are difficult to solve. 
However, if one is looking for exact solutions admitting a space-time symmetry, one may proceed via the so called mini-superspace approach (see for example Refs.~\cite{Vil, capozziello,
monica}).

In this paper we consider a class of static spherically symmetric topological space-times defined by the metric
\begin{equation}
 ds^2=-\text{e}^{2\alpha(r)}B(r)+\frac{dr^2}{B(r)}+r^2d\Omega^2_k\,,
 \label{metric}
\end{equation}
where $d\Omega^2_k$ is the metric of a constant curvature compact two-dimensional space, the so called horizon manifold with areal radius $r$, and admitting  
three different topologies, namely spherical, flat (toroidal really) or Riemann surfaces, depending on the $k$ parameter, $k=1\,,0\,,-1$, respectively. 
Furthermore, $\alpha(r)$ and $B(r)$ are functions of the radial coordinate only.

The associated Ricci scalar reads,
\begin{eqnarray}
R  &=&
-3\,\left[{\frac{d}{dr}}B\left(r\right)\right]{\frac{d}{dr}}
\alpha\left(r\right)-2\,B\left(r\right)\left[{\frac{d}{dr}}
\alpha\left(r\right)\right]^{2}-{\frac{d^{2}}{d{r}^{2}}}
B\left(r\right)-2\,B\left(r\right){\frac{d^{2}}{d{r}^{2}}}\alpha\left(r\right)\nonumber\\
&&-4\,{\frac{{\frac{d}{dr}}B\left(r\right)}{r}}
-4\,{\frac{B\left(r\right){\frac{d}{dr}}\alpha\left(r\right)}{r}}-2\,{\frac{B\left(r\right)}{{r}^{2}}}
+\frac{2k}{{r}^{2}}\,.\label{R}
\end{eqnarray}
In what follows, we implement the mini-superspace approach following Ref.~\cite{Seba}.

First from the Appendix A we note that in the case of the metric (\ref{metric})
the related boundary term is a total divergence with respect to $r$ and  may be written as
\begin{eqnarray}
BT &=&-V_k \int dt \int dr \frac{d}{dr}\left[
F'(R)\text{e}^{\alpha(r)}r^2\left(\frac{d B(r)}{d r}+2 B(r)\frac{d\alpha(r)}{d r}+\frac{4 B(r)}{r}\right)
\right]\,,\nonumber\\
&=&
-V_k \int  dt \left[
F'(R)\text{e}^{\alpha(r)}r^2\left(\frac{d B(r)}{d r}+2 B(r)\frac{d\alpha(r)}{d r}+\frac{4 B(r)}{r}\right)
\right]\,,
\label{BT}
\end{eqnarray}
where $V_k$ is the volume of the horizon manifold, namely $V_1 = 4\pi$ for the sphere,
$V_0 =\mathrm{Im}\tau$, with $\tau$ the Teichm\"{u}ller parameter for the torus, and finally $V_{-1} = 4\pi(g-1)$, $2<g$, for the compact hyperbolic manifold with genus $g$~\cite{genus}.

In order to deal with a standard Lagrangian with quantities admitting only first order derivatives
with respect to $r$, one may introduce in the action (\ref{action0}) evaluated with respect to the metric (\ref{metric}) a Lagrangian multiplier $\lambda$ in the following way,
\begin{eqnarray}
  I&=&\int_\mathcal{M} d^4 x\left(\text{e}^{\alpha(r)}r^2\right)\left[F(R)-\lambda\left[R
  +3\,\left[{\frac{d}{dr}}B\left(r\right)\right]{\frac{d}{dr}}
\alpha\left(r\right)+2\,B\left(r\right)\left[{\frac{d}{dr}}
\alpha\left(r\right)\right]^{2}
\right.\right.\nonumber\\&&\hspace{-0.5cm}\left.
\left.
+{\frac{d^{2}}{d{r}^{2}}}
B\left(r\right)+2\,B\left(r\right){\frac{d^{2}}{d{r}^{2}}}\alpha\left(r\right)
+4\,{\frac{{\frac{d}{dr}}B\left(r\right)}{r}}
+4\,{\frac{B\left(r\right){\frac{d}{dr}}\alpha\left(r\right)}{r}}+2\,{\frac{B\left(r\right)}{{r}^{2}}}
-\frac{2k}{{r}^{2}} \right]\right]\,.
\end{eqnarray}
Thus, the variation with respect to $R$ leads to the equation (\ref{R}) after the identification
\begin{equation}
\lambda=F'(R)\,. \label{lambda}
\end{equation}
Now, integrating by parts, it is possible to write the action in the standard form with respect to the variables $\alpha(r)\,,B(r)$ and $R=R(r)$, 
namely
\begin{eqnarray}
I &=&V_k\int dt\,\int dr e^{\alpha(r)}\left\{r^2\left(F(R)-F'(R)R\right)+F'(R)\left(2k+2 r\frac{d B(r)}{dr}+2 B(r)+4r B(r)\frac{d\alpha(r)}
{d r}\right)\right.
\nonumber\\
&& +\left.F''(R)\frac{d R}{d r}r^2\left(\frac{d B(r)}{d r}+2B(r)\frac{d \alpha(r)}{dr}+\frac{4B(r)}{r}\right)\right\}+BT\,,
\label{A}
\end{eqnarray}
where we take into account the equalities in (\ref{BT}) and (\ref{lambda}). 
As a consequence, one may work only with the new bulk action, obtained subtracting the correct boundary term,
\begin{eqnarray}
\hat{I} &=&V_k\int dt\,\int dr e^{\alpha(r)}\left\{r^2\left(F(R)-F'(R)R\right)+F'(R)\left(2k+2 r\frac{d B(r)}{dr}+2 B(r)+4r B(r)\frac{d\alpha(r)}
{d r}\right)\right.
\nonumber\\
&& +\left.F''(R)\frac{d R}{d r}r^2\left(\frac{d B(r)}{d r}+2B(r)\frac{d \alpha(r)}{dr}+\frac{4B(r)}{r}\right)\right\}\,.
\label{NA}
\end{eqnarray}
Finally, the equations of motion can be obtained by making the variation with respect $\alpha(r)$ and $B(r)$ and are given by (see also Appendix B),
\begin{eqnarray}
& &V_k\text{e}^\alpha(r)\left[r^2\left(RF'(R)-F(R)\right)-2F'(R)\left(k-B(r)-r\frac{d B(r)}{d r}\right)\right.\nonumber\\
& & \left.+2B(r)F''(R)r^2\left[\frac{d^2 R}{d
r^2}+\left(\frac{2}{r}+\frac{dB(r)/dr}{2 B(r)}\right)\frac{d R }{d
r}+\frac{F'''(R)}{F''(R)}\left(\frac{d R}{d
r}\right)^2\right]\right]=0\,,\label{one}
\end{eqnarray}
\begin{equation}
V_k\text{e}^\alpha(r)\left[\frac{1}{r^2}\frac{d\alpha(r)}{dr}\left(\frac{2}{r}+\frac{F''(R)}{F'(R)}\frac{d
R}{d r}\right)-\frac{1}{r^2}\frac{F''(R)}{F'(R)}\frac{d^2 R}{d
r^2}-\frac{1}{r^2}\frac{F'''(R)}{F'(R)}\left(\frac{d R}{d
r}\right)^2\right]=0\,.\label{two}
\end{equation}
Furthermore, as  already mentioned, 
the variation with respect to $R$ leads again to Eq.~(\ref{R}). With this approach, Eq.~(\ref{one}) does not contain an explicit (non trivial) dependence on $\alpha(r)$, while Eq.~(\ref{two}) 
does not contain an explicit dependence on $B(r)$.
In order to look for exact solutions, 
the strategy is to make suitable Ansatz for $R=R(r)$ or to make an Ansatz for $\alpha(r)$. In the next subsections, we will review the  examples of BH solutions we are interested in.

\subsection{Constant curvature case\label{s1}}

In the constant Ricci scalar case one has $R=R_0$. From Eq.~(\ref{two}) we immediately obtain
\begin{equation}
\alpha(r)=\text{const}. 
\end{equation}
Thus, if $F'(R_0)\neq 0$, Eq.~(\ref{one}) leads to the topological Schwarzschild-AdS solution\footnote{Since in this paper we are interested in black hole solutions with a well 
defined temperature
we will not consider de Sitter metrics with two horizons.},
\begin{equation}
B(r)=k-\frac{c}{r}-\frac{\Lambda r^2}{3}\,,\quad
\Lambda=\frac{R_0 F'(R_0)-F(R_0)}{2 F'(R_0)}<0\,,\label{SdS}
\end{equation}
where $c$ is a free integration constant. Finally,
from Eq.~(\ref{R}) one has,
\begin{equation}
R_0=4\Lambda\,,
\end{equation}
such that $\Lambda=F(R_0)/(2F'(R_0))$.

\subsection{Solutions with $\alpha(r)=\text{const}$\label{s2}}

The Equation (\ref{two}) with $\alpha(r)=\text{const}$ leads to~\cite{iran, Seba},
\begin{equation}
F'(R)=a r+b \,, \label{An2}
\end{equation}
where $a\,,b$ are constant parameters. The form of $B(r)$ 
can be derived by taking the derivative respect to $r$ of the Equation (\ref{one}), but in general
it is not possible to fully  reconstruct the corresponding $F(R)$-model (see Refs.~\cite{Seba, cogno} for details).
On the other hand, when $a=0$ we recover the constant Ricci scalar case already treated in the preceding subsection, while if one poses $b=0$ 
we get
\begin{equation}
 B(r)=\frac{k}{2}+\frac{c}{r^2}+\lambda r^2\,, \label{BF1}
\end{equation}
where $c\,,\lambda$ are integration constants. The Ricci scalar reads,
\begin{equation}
 R=-12\lambda +\frac{k}{r^2}\,,
\end{equation}
and by using Eq.~(\ref{An2}) one easily reconstruct the model as
\begin{equation}
F(R)=2a k\sqrt{k(R+12\lambda)}\,. \label{F1}
\end{equation}
Note that in this case only one free integration constant $c$ appears in the metric.

\subsection{Clifton-Barrow solutions\label{s3}}

Consider the Lagrangian 
\begin{equation}
F(R)=\frac{R^{\delta+1}}{\kappa}\,,\quad \delta\neq 1\,,\label{CFmodel} 
\end{equation}
with $\kappa$ a dimensional parameter. One looks for solutions described by the SSS metrics with $\alpha(r)\neq 0$, namely
\begin{equation}
\text{e}^{2\alpha(r)}=\left(
\frac{r}{r_0}
\right)^{2a}\,,\label{z}
\end{equation}
where $a$ is a number and $r_0$ a dimensional constant. We  also assume 
\begin{equation}
 R=\frac{R_0}{r^2}\,.
\end{equation}
In this case, Eqs.~(\ref{one})--(\ref{two}) are solved by ( $k=1$ in Ref.~\cite{CB}, $k$ generic in Ref.~\cite{Gorbu}),
\begin{equation}
\alpha(r)=\log\left[
\left(\frac{r}{r_0}\right)^a
\right]\,,\quad
B(r)=B_0\left(k-\frac{c}{r^b}\right) \,,
\end{equation}
where $c$ is a free integration constant and 
$R_0\,, B_0\,, a\,, b$ are functions of the parameter $\delta$,
\begin{eqnarray}
& & R_0=\frac{6\delta k(1+\delta)}{(2\delta^2+2\delta-1)}\,,\quad
B_0=\frac{(1-\delta)^2}{(1-2\delta+4\delta^2)(1-2\delta-2\delta^2)}\,,\nonumber\\ & &
a=\frac{\delta(1+2\delta)}{(1-\delta)}\,,\quad
b=\frac{(1-2\delta+4\delta^2)}{(1-\delta)}\,.\label{CFrel}
\end{eqnarray}
We also observe that the following relation holds true:
\begin{equation}
b=a-2\delta+1\,. \label{condCB}
\end{equation}
When $\delta=-1/2$ one has $\alpha(r)=\text{const}$ and we recover the model (\ref{F1}) with $\lambda=0$ and solution (\ref{BF1}).
The case $\delta=1$ has to be considered separately and corresponds to the scale invariant model $F(R)\sim R^2$ (see for example \cite{max}) and will not be investigated in this paper.

It is also possible to add to the Clifton-Barrow model in (\ref{CFmodel}) a cosmological constant.
An explicit example is the following:  $a=2$, thus $\delta=-2$. The corresponding  model with cosmological constant is given by~\cite{cogno},
\begin{equation}
F(R)=\frac{1}{\kappa}\left(\frac{1}{R}-\lambda\right)\,.\label{mSeba} 
\end{equation}
 When $k \neq 0$, the model admits the 
topological SSS solution (\ref{metric}) with
\begin{equation}
 \text{e}^{2\alpha(r)}=\left(
\frac{r}{r_0}
\right)^{4}\,,\quad B(r)=-\frac{k}{7}+\frac{c}{r^7}+\frac{8\lambda}{15 r^2}\,,\quad R=\frac{4k}{r^2}\,.
\end{equation}
Furthermoe, when $\lambda\neq 0$ the model in (\ref{mSeba}), after the redefinition $\lambda\rightarrow 6k/\lambda^2$,  
leads to the solution~\cite{Seba}:
\begin{equation}
  \text{e}^{2\alpha(r)}=\left(
\frac{r}{r_0}
\right)\,,\quad
B(r)=\frac{4}{7}
\left(
k+\frac{c}{r^{7/2}}-\frac{7\lambda r}{36}
\right)\,,\quad R=\frac{\lambda}{r}\,.
\end{equation}

\section{First law and BH energy in $F(R)$-gravity}

In this Section, following Ref.~\cite{Gorbu}, we propose a simple method to obtain the black hole energy in $F(R)$-gravity
by starting from the First Law of Thermodynamics (see also Refs.~\cite{Deserenergy,Abreu,Cai}).

We recall that a SSS solution in the form of (\ref{metric})  describes a  black hole  
with a single  event horizon with radius $r=r_H$ when there exists a single $r_H>0$ such that
\begin{equation}
B(r_H)=0\,,\quad 0<\frac{d B(r)}{d r}|_{r=r_H}\,.\label{BHcond}
\end{equation}
 In this way, $0<d B(r)/d r|_{r_H}$ leads to a positive Killing surface gravity
\begin{equation}
\kappa_K=\text{e}^{\alpha(r_H)}\frac{d B(r)}{d r}|_{r=r_H}\,.
\end{equation}
The metric signature $(- + + +)$ 
is preserved for $r_H<r$, while is violated when $r<r_H$. In other words, inside of the horizon, the coordinate $r$ plays the role of the time and $t$ plays the role of a spatial coordinate, and the  metric becomes dynamic.

The metrics presented in the preceding section describe a black hole at least for some choices of the horizon topology. For example, it is well known that the Schwarzschild-AdS metric in (\ref{SdS}) 
can describe a black hole with various topologies when $\Lambda < 0$
(see Ref.~\cite{Vanzo}), but if $\Lambda=0$ we obtain a black hole only for $k=1$ and $0<c$.

Given a BH solution within a $F(R)$-modified gravity model, it is well known that the  entropy and the related Hawking tempertaure can be computed by making use of  independent approaches. 
In the case of the black hole entropy, Wald method gives~\cite{Wald},
\begin{equation}
 S_W=(4\pi)V_k r_H^2 F'(R_H)\,, 
\end{equation}
where the pedex $H$ denotes a quantity evaluated with respect to $r=r_H$.
The Killing-Hawking temperature~\cite{HT} can be derived, for instance, with the tunneling method~\cite{parikh,noi} and reads,
\begin{equation}
T_K=\frac{\kappa_K}{2\pi}\equiv\frac{\text{e}^{\alpha(r_H)}}{4\pi}\frac{d B(r_H)}{d r}\,. 
\end{equation}
Thus, from Eq.~(\ref{one}) evaluated on the BH horizon we may derive a First Law of Thermodynamics where the Killing temperature emerges in a natural way as follows,
\begin{equation}
T_k d S_W
=\text{e}^{\alpha(r_H)} V_k\left(2 k\,F'(R_H)-\left(R_H F'(R_H)-F(R_H)r_H^2\right)\right)dr_H\,.
\end{equation}
Here, we have used the condition $B(r_H)=0$ and we have multiplied by $dr_H$. An important remark is in order. The relation
\begin{equation*}
d S_W=\left(4\pi\right)V_k\left(2 r_H F'(R_H)dr_H+r_H^2 F''(R_H)\left(\frac{d R}{dr}\right)\Big\vert_H dr_H\right)\,,
\end{equation*}
is valid only if 
\begin{equation}
d R_H= \left(\frac{d R}{dr}\right)\Big\vert_H dr_H\,.\label{condS}
\end{equation}
It means that the on shell form of the Ricci scalar does not have to depend on the integration constant(s) of the solution. In this case the First Law holds true,
 \begin{equation}
T_K d S_W= d E_K\,,\label{FirstLaw0}
\end{equation}
and leads to the identification
\begin{equation}
d E_K= \text{e}^{\alpha(r_H)} V_k\left(2 k\,F'(R_H)-\left(R_H F'(R_H)-F(R_H)r_H^2\right)\right)d r_H\,.
\end{equation}
Thus, at least in the case where only an integration constant appears in the black hole solution, we have an explicit expression for the BH energy in $F(R)$-gravity,
\begin{equation}
E_K:=V_k\int  \text{e}^{\alpha(r_H)}\left(2 k\,F'(R_H)-\left(R_H F'(R_H)-F(R_H)r_H^2\right)\right)d r_H\,.
\label{FirstLaw}
\end{equation}
The condition (\ref{condS}) looks restrictive, but it holds for a large class of static black holes in $F(R)$-gravity. For these
solutions,  the First Law is a robust argument for the definition of the BH energy or mass.
In particular, these considerations are valid for the $F(R)$-models presented in the previous section.

\section{The evaluation of the action growth in $F(R)$-gravity}

In this Section, we start recalling the approach described in Ref.~\cite{Susskind} and used 
within $F(R)$ gravity in other papers (see for example Ref.~\cite{Ali, Wa}). In Ref.~\cite{Myers} this approach has been rigorously proved to give the correct answer. 
Within an holographic scenario,  the complexity-action conjecture (CA) tells us that one can compute the complexity
growth by the evaluation of the action growth in the time, action defined with respect to the so called Wheeler-de Witt
(WdW) patch. For large time, one may consider only the bulk on-shell action associated with the black hole solution evaluted in the interior region of the black hole.

Motivated by these considerations, we shall compute the action growth in $F(R)$-modified 
gravity making use of our mini-superspace approach 
and working only with the bulk action (\ref{NA}). We are interested in computing 
the action growth associated with the interior of the black holes, namely for $r<r_H$, where the metric becomes dynamics. 
Thus, if we replace $r$ with a time coordinate and $t$ with a space coordinate, 
\begin{equation}
r=T\,,\quad t=\rho\,,\quad 0<T<r_H\,, 
\end{equation}
we can see the interior metric as a Spherically Symmetric Dynamical (SSD) space-time. The metric (\ref{metric}), after the redefinition $B(r)\rightarrow -B(T)$,  
can be rewritten as,
\begin{equation}
 ds^2=-\frac{dT^2}{B(T)}+\text{e}^{2\alpha(T)}B(T)d\rho^2+ T^2d\Omega^2_k=\gamma_{ij}(x^i)dx^idx^j+(\mathcal R)^2d\Omega^2_k\,,
 \label{imetric}
\end{equation} 
where $\gamma_{ij}(x^i)$ is the reduced metric with respect to the coordinates $x^i=(T\,,\rho)$, $\mathcal R(x^i)=T$ is the areal radius,
and $B(T)$ is positive in the given range of $T$. In a dynamical case we lose
the time-like Killing vector field and the Killing formalism becomes meaningless. On the other hand, 
one can use the covariant Hayward formalism \cite{Hay}. The trapping (event horizon) is located at
\begin{equation}
\chi=\gamma^{ij}\partial_i R(x^i)\partial_j R(x^i)=0\,,
\end{equation}
and one has $B(T_H)=0$. Furthermore, Hayward surface gravity is
\begin{equation}
\kappa_H=\frac{1}{2}\Box_\gamma\mathcal R(x^i)_H\,,  
\end{equation}
where the d'Alambertian is referred to the reduced metric.
In our case 
\begin{equation}
 \kappa_H=-\frac{1}{2}\frac{d B(T)}{d T}|_{T=T_H}\,.
\end{equation}
Here,  the role of the time-like Killing
vector is played by the  Kodama vector~\cite{Kodama},
\begin{equation}
  K^i=\frac{1}{\sqrt{-\gamma}}\varepsilon^{ij} \partial_j R(x^i)\,,
\end{equation}
where $\gamma$ is the determinant of the reduced metric $\gamma_{ij}(x^i)$ and $\epsilon^{ij}$ is the two-dimensional antisymmetric Levi-Civita tensor.
Thus,  we get 
\begin{equation}
K^\mu=(0,\text{e}^{-\alpha(T)},0,0)\,.
\end{equation}
The action growth can be defined in a covariant way by means
\begin{equation}
 C= \lim_{T \rightarrow r_H} K^\mu \partial_\mu \hat{I}\,,
 \label{can}
\end{equation} 
where $\hat I$ is the bulk action (\ref{NA}).

For our class of black hole solutions within the modified gravitational theories described by $F(R)$, one has
\begin{equation}
 C=V_k \text{e}^{-\alpha(r_H)}\int_0^{r_H} dT L(T) \,,
 \label{ca2}
\end{equation}  
where the bulk Lagrangian is given by
\begin{eqnarray}
L(T) &=& e^{\alpha(T)}\left\{T^2\left(F(R)-F'(R)R\right)+2F'(R)\left(k- T\frac{d B(T)}{dT}- B(T)-2T B(T)\frac{d\alpha(T)}
{d T}\right)\right.
\nonumber\\
&& -\left. F''(R)\frac{d R}{d T}T^2\left(\frac{d B(T)}{d T}+2B(T)\frac{d \alpha(T)}{dT}+\frac{4B(T)}{T}\right)\right\}\,,
\label{call}
\end{eqnarray}
and must be evaluated on shell, $R=R(T)$ being a function of $T$, 
\begin{eqnarray}
R  &=&
3\,\left[{\frac{d}{dT}}B\left(T\right)\right]{\frac{d}{dT}}
\alpha\left(T\right)+2\,B\left(T\right)\left[{\frac{d}{dT}}
\alpha\left(T\right)\right]^{2}+{\frac{d^{2}}{d{T}^{2}}}
B\left(T\right)+2\,B\left(T\right){\frac{d^{2}}{d{T}^{2}}}\alpha\left(T\right)\nonumber\\
&&+4\,{\frac{{\frac{d}{dT}}B\left(T\right)}{T}}
+4\,{\frac{B\left(T\right){\frac{d}{dT}}\alpha\left(T\right)}{T}}+2\,{\frac{B\left(T\right)}{{T}^{2}}}
+\frac{2k}{{T}^{2}}\,.\label{RT}
\end{eqnarray}

\subsection{Action growth: $\alpha=0$ cases.}
 
In this subsection, we will calculate the action growth of the $F(R)$-black holes with metric (\ref{metric}) and $\alpha(r)=\text{const}$. 
Without loss 
of generality we can pose $\alpha(r)=0$.
 Let us start with the constant Ricci curvature case $R=R_0$ analyzed in \S\ref{s1}. 
Evaluating the Lagrangian (\ref{call}) on the solution,
\begin{equation}
B(T)=-k+\frac{c}{T}+\frac{\Lambda T^2}{3}\,,\quad
\Lambda=\frac{R_0 F'(R_0)-F(R_0)}{2 F'(R_0)}\,,
\end{equation}
one has 
\begin{equation}
L(T)=4 F'(R_0)(k-\Lambda T^2)\,.
\end{equation}
As a result we obtain for the action growth (\ref{ca2}),
\begin{equation}
C= 4V_{k}F'(R)\left(k r_H-\frac{1}{3}\Lambda r_H^3\right)\,.
\end{equation}
By using the horizon condition $B(T_H)=0$,  we get
\begin{equation}
C=4 V_{k}F'(R)c\,.
\end{equation}
In general, from (\ref{FirstLaw}), we can now identify
\begin{equation}
E_K=2V_k F'(R_H)c\,. \label{ESdS}
\end{equation}
As a result one gets,
\begin{equation}
C= 2E_K\,.\label{C2EK}
\end{equation}
This result is in agreement with the action growth computed by other method in Refs.~\cite{Caibound, FRbound, Ali}. 
One remark is in order. The expression (\ref{ESdS}) has been obtained by fixing the cosmological constant $\Lambda$ and 
the procedure is always valid when $\Lambda$ explicitly appears in the form of the $F(R)$-model (for example, $F(R)\propto R-2\Lambda$). However, 
when the cosmological constant is a second 
integration constant of the solution, an additional thermodynamical potential may contribute to the energy. It is the case, for instance, of
$R^2$-gravity, where the scale invariance of the theory brings to the emergence of the lenght scale from the solution.

Let us come back to the model (\ref{F1}) discussed in \S\ref{s2}, for which one has $\alpha(T)=0$, but non trivial Ricci curvature.
We have that the model admits the following interior BH solution,
\begin{equation}
 B(T)=-\frac{k}{2}-\frac{c}{T^2}-\lambda T^2\,, 
\end{equation}
with non-constant Ricci scalar,
\begin{equation}
 R=-12\lambda +\frac{k}{T^2}\,.
\end{equation}
The on-shell Lagrangian is
\begin{equation}
 L(T)=6 a T(k+4\lambda T^2)\,.
\end{equation}
Thus, the action growth results to be,
\begin{equation}
 C=6 V_k a\left(\frac{k r_H^2}{2}+\lambda r_H^4\right)\equiv -6 V_K a c\,.
\end{equation}
From (\ref{FirstLaw}) we have that the energy of the black hole under investigation is,
\begin{equation}
 E_K=-3 V_K a c\,,
\end{equation}
and one obtains again the relation (\ref{C2EK}).
This is a new result, similar to the the case with constant Ricci scalar discussed above.

\subsection{Action growth: Clifton-Barrow models}

As an example of non-constant Ricci scalar case with $\alpha(r)\neq 0$, we compute the action growth for
the Clifton-Barrow models (\ref{CFmodel}) discussed in \S\ref{s3}.
The interior BH solution reads,
\begin{equation}
\alpha(T)=\log\left[\left(\frac{T}{r_0}\right)^a\right]\,,\quad
B(T)=B_0\left(-k+\frac{c}{T^b}\right) \,,\quad R=\frac{R_0}{T^2}\,,
\end{equation}
where $R_0\,,B_0\,,a$ and $b$ are given by (\ref{CFrel}).
By using the definitions in (\ref{ca2})--(\ref{call}) and the condition (\ref{condCB}), the action growth results to be
\begin{equation}
C=
\frac{V_k e^{-\alpha(T)}}{\kappa r_0^a}R_0^\delta B_0(1-\delta^2)k r_H^b\equiv
\frac{V_k e^{-\alpha(T)}}{\kappa r_0^a}R_0^\delta B_0(1-\delta^2)  c
\,.
\end{equation}
On the other hand, the Killing BH energy for a Clifton-Barrow BH is derived as
\begin{equation}
E_K=2(1-\delta^2) \frac{V_k}{r_0^a\kappa}R_0^\delta B_0 c\,,
\end{equation}
and, as a consequence, one  has 
\begin{equation}
 C=2 \text{e}^{-\alpha(r_H)} E_K\,.\label{u}
\end{equation}
We will return later on this result.

As a further example, we will consider now the model (\ref{mSeba}) with interior BH solution,
\begin{equation}
 \text{e}^{2\alpha(T)}=\left(
\frac{T}{r_0}
\right)^{4}\,,\quad B(T)=\frac{k}{7}-\frac{c}{T^7}-\frac{8\lambda}{15 T^2}\,,\quad R=\frac{4k}{T^2}\,,
\end{equation}
where we recall that $k\neq 0$. For the action growth one obtains,
\begin{equation}
C=\frac{3V_k\text{e}^{-\alpha(r_H)}}{4r_0^2\kappa}\left(\frac{k r_H^7}{7}-
\frac{8\lambda r_H^5}{15}\right)
\equiv\frac{3 \text{e}^{-\alpha(r_H)}c}{4r_0^2 \kappa}\,. 
\end{equation}
Since the BH Killing energy computed with the static external metric reads,
\begin{equation}
E_K= \frac{3 c}{8r_0^2 \kappa},
\end{equation}
we see that the relation (\ref{u}) holds true again. The result is confirmed 
even in the case of the interior BH solution 
\begin{equation}
  \text{e}^{2\alpha(T)}=\left(
\frac{T}{r_0}
\right)\,,\quad
B(T)=-\frac{4}{7}
\left(
k+\frac{c}{T^{7/2}}-\frac{7\lambda T}{36}
\right)\,,\quad R=\frac{\lambda}{T}\,,
\end{equation}
which can be still inferred from the model (\ref{mSeba}) after the redefinition 
$\lambda\rightarrow 6k/\lambda^2$. Now the action growth is
\begin{equation}
C=\frac{8 V_k \text{e}^{-\alpha(r_H)}}{63\lambda^2
\kappa(r_0)^{1/2}}r_H^{7/2}\left(-36k+7\lambda r_H\right)\equiv
\frac{32V_K \text{e}^{-\alpha(r_H)} c}{7\kappa\lambda^2(r_0)^{1/2}}\,,
\end{equation}
while the Killing energy of the black hole is derived as
\begin{equation}
E_K=\frac{16V_K c}{7\kappa\lambda^2(r_0)^{1/2}}\,.
\end{equation}
It follows that for these classes of black hole solutions with $\alpha(r)\neq 0$, the action growth has the universal form (\ref{u}),
namely the Kodama-Hayward energy,
\begin{equation}
E_H=\text{e}^{-\alpha(r_H)} E_K\,,
\end{equation}
appears. When $\alpha=0$ we recover the relation (\ref{C2EK}).

\section{Deser-Sarioglu-Tekin black holes}

So far, we have investigated in detail the BH solution within $F(R)$-modified gravity, and we have obtained a quite general result expressed by (\ref{u}). 
In this Section we would like to present another specific example of modified gravity for which we can make use of the mini-superspace approach.

In Ref.~\cite{deser}, Deser, Sarioglu and Tekin presented an interesting model based on a Weyl correction to GR 
for which they provide an exact SSS BH solution.
The model including the cosmological constant has the following action,
\begin{equation}
I = \frac{1}{2\kappa^2} \int_{\mathcal M}\,d^4x\,\sqrt{-g} 
\left(R -2\Lambda + \sqrt{3}\sigma\,\sqrt{W}\right) \,, \label{d action}
\end{equation}
\phantom{line}\\
where $\Lambda$ is the cosmological constant, $\sigma$ is a real dimensionless parameter 
and $W=C_{\mu\nu\xi\sigma}C^{\mu\nu\xi\sigma}$ is the square of the Weyl tensor,
\begin{equation}
W=\frac{1}{3}R^{2}-2R_{\mu\nu}R^{\mu\nu}+R_{\mu\nu\xi\sigma}R^{\mu\nu\xi\sigma}\,,\label{Weylsquare} 
\end{equation}
$R_{\mu\nu}$ and $R_{\mu\nu\sigma\xi}$ being the Ricci and the Riemann tensors, respectively.
For $\sigma=0$ the Weyl contribution turns off and the action of $\Lambda$CDM Model is recovered for $\kappa^2=16\pi G_N$, with $G_N$ the Newton constant.
A key point is the following: for the SSS metric (\ref{metric}) the square of the Weyl tensor is a perfect square and reads,
\begin{eqnarray}
W&=& \frac{1}{3} \left[ \frac{1}{r^2}
\left[r^2 \left(\frac{d^2 B(r)}{d r^2}\right) + 2\left(B(r)-k\right) -2 r \left(\frac{d B(r)}{d r}\right)\right]\right.\nonumber\\ 
&&\left.\hspace{-2cm} + \frac{1}{r} \left[3 r \left(\frac{d B(r)}{d r}\right)\left(\frac{d\alpha(r)}{d r}\right) -2 B(r) \left(\frac{d \alpha(r)}{d r}
-r\left(\frac{d^2 \alpha(r)}{d r^2}+\left(\frac{d\alpha(r)}{d r}\right)^2\right)\right)\right]\right]^2\,.\label{C2} 
\end{eqnarray}
After integration by parts, we are able to separate the action of the bulk from the boundary terms (see Appendix C) as 
\begin{eqnarray}
 I&=&\frac{V_k}{2\kappa^2}\int dt\int dr
 \text{e}^{\alpha(r)}\left(
 -2\Lambda r^2+2k(1-\epsilon\sigma)+
 2B(r)(1-\epsilon\sigma)+2r \frac{d B(r)}{d r}(1-4\epsilon\sigma)\right.\nonumber\\&&\left.
 +2rB(r)\frac{d\alpha(r)}{d r}(2-5\epsilon\sigma)
 \right)+BT\,,
\end{eqnarray}
where
\begin{eqnarray}
BT &=&-\frac{V_k}{2\kappa^2}\int dt\int dr\frac{d}{dr}\left[
\text{e}^{\alpha(r)}r^2\left(\frac{d B(r)}{d r}+2B(r)\frac{d\alpha(r)}{d r}+\frac{4B(r)}{r}\right)\left(
1-\epsilon\sigma
\right)
\right]\nonumber\\
&=&
-\frac{V_k\Delta t}{2\kappa^2}\left[
\text{e}^{\alpha(r)}r^2\left(\frac{d B(r)}{d r}+2B(r)\frac{d\alpha(r)}{d r}+\frac{4B(r)}{r}\right)\left(
1-\epsilon\sigma
\right)
\right]
\,.\label{BTWeyl}
\end{eqnarray}
In this
expression, the parameter $\epsilon=\pm 1$ must be set in order to have
$\sqrt{W}=|\sqrt{W}|$. The bulk action is obtained by subracting the boundary term, namely
\begin{eqnarray}
 \hat I&=&\frac{V_k}{2\kappa^2}\int dt\int dr
 \text{e}^{\alpha(r)}\left(
 -2\Lambda r^2+2k(1-\epsilon\sigma)+
 2B(r)(1-\epsilon\sigma)+2r \frac{d B(r)}{d r}(1-4\epsilon\sigma)\right.\nonumber\\&&\left.
 +2rB(r)\frac{d\alpha(r)}{d r}(2-5\epsilon\sigma)
 \right)\,.\label{Cb}
\end{eqnarray}
The field equations are derived by making the variation of the bulk action with respect to $\alpha(r)$ and $B(r)$ and read,
\begin{equation}
\frac{V_k}{2\kappa}\text{e}^{\alpha(r)}\left[(1-\epsilon\,\sigma)\left(k-B(r)-r \frac{d B(r)}{d r}\right)
+ 3 \epsilon\sigma B(r)-\Lambda r^2\right]= 0\,, 
\label{EOM1Deser}
\end{equation}
\begin{equation}
\frac{V_k}{2\kappa}\text{e}^{\alpha(r)}\left[3\epsilon\sigma+\frac{d\alpha(r)}{d r}(1-\epsilon\sigma)r\right]=0\,.
\label{EOM2Deser}
\end{equation}
Note that for the SSS metric the field equations of this theory are at the second order.
The general solution is given by~\cite{Gorbu, deser},
\begin{equation}
 \alpha(r)= \log\left[\frac{r}{r_0}\right]^{\frac{3\epsilon\sigma}{\epsilon\sigma-1}},
\quad
B(r)= k\,\frac{(1-\epsilon\sigma)}{(1-4\epsilon\sigma)} - c r^{-\frac{1-4\epsilon\sigma}{1-\epsilon\sigma}}
 -\Lambda\,\frac{r^2}{3(1-2\epsilon\sigma)}\,,\quad\sigma\neq \pm1\,,\pm\frac{1}{4}\,,\label{b}
\end{equation}
where $r_0$ has been introduced for dimensional reasons, and $c$ is an integration constant.
This solution describes a black hole with event horizon located at $B(r_H)=0$.  Thus, if one uses the Killing temperature and the Wald entropy~\cite{Bellini},
\begin{equation}
T_K=\frac{1}{4\pi}\left(\frac{r_H}{r_0}\right)^{\frac{3\epsilon\sigma}{\epsilon\sigma-1}}\left(
c\left(\frac{1-4\epsilon\sigma}{1-\epsilon\sigma}\right)r_H^{-\frac{-3\epsilon\sigma}{1-\epsilon\sigma}}-2\Lambda\,\frac{r_H}{3(1-2\epsilon\sigma)}
\right)\,,\quad S_W=(4\pi )\frac{V_k r_H^2}{2\kappa}\left(1-\epsilon\sigma\right)\,,
\end{equation}
it is easy to see that Eq.~(\ref{EOM1Deser}) evaluated on the horizon leads to  the First Law of Thermodynamics, namely
\begin{equation}
T_K d S_W=\frac{V_k\text{e}^{\alpha(r_H)}}{\kappa}\left[
(1-\epsilon\sigma)k-\Lambda r_H^2\right]\equiv d E_K\,.
\end{equation}
Thus, we can identify the BH energy as,
\begin{equation}
E_K: =
\frac{V_k}{\kappa}
\int
\text{e}^{\alpha(r_H)}\left[
(1-\epsilon\sigma)k-\Lambda r_H^2\right]d r_H
=\frac{V_k c}{\kappa r_0^{\frac{3\epsilon\sigma}{\epsilon\sigma-1}}}(1-\epsilon\sigma)\,,
\end{equation}
where we have taken into account that  on the horizon  $B(r_H)=0$.

The growth action can be computed in an analogue way of the $F(R)$-case by starting from (\ref{Cb}) and the result is
\begin{equation}
C=\text{e}^{-\alpha(r_H)}E_K\frac{(2-5\epsilon\sigma)}{(1-\epsilon\sigma)}\leq 2\text{e}^{-\alpha(r_H)}E_K\,.
\end{equation}
In this case of modified gravity model, the action growth does not coincide with the double of the Kodama energy, but is still proportional
to it and, more importantly, bounded by twice its value as long as $0<\epsilon\sigma$, which is in accord with the general complexity bound as usually stated.  In the contrary case $\epsilon\sigma<0$ and the bound is violated. 
As a check, when $\sigma $ goes to zero, one gets the result of General Relativity.

\section{A bit of black hole phenomenology}

What we say in this section is strictly valid in Einstein's theory of gravity and then argued to hold for more general models. 
Black hole radiates aways their mass in a certain lifetime. The efficiency of particle emission from black holes is clearly an important issue of the evaporation phenomenon.
Beyond this, it is also relevant to interpret the result on the action grow that we obtained in some models of modified gravity. In particular we would like to justify the use of non rotating uncharged solutions.  

As is well known, the temperature of a  black hole in GR is inversely proportional to its total mass, $M$, which includes the gravitational contribution, and the horizon 
area $A$ is proportional to $M^2$, so the total power emitted $P$ is proportional 
to $AT^4$, or $M^{-2}$. From this it follows that the lifetime $t_l=M/P$ is 
proportional to $M^3$. To state a number,  black hole formed by 
stellar collapse having $ M_{\odot}\leq M$, where $M_{\odot}$ is the solar mass, have a lifetime of 
order $10^{66}$ yr. Therefore the thermal emission is physically  insignificant 
for such black holes, although still very important theoretically. However, it is relevant for primordial black holes, for which $M$ could be less than $10^{15}$ g, and the corresponding lifetime 
less than the age of the universe.

The problem is how rapidly a charged rotating black hole 
discharges and spins down. The main difference in the rapidity of the 
two processes can be seen as follows. Given the 
BH angular momentum $J$
and the BH charge $Q$,
the two parameters 
\begin{equation}
a_*=\frac{J}{M^2}, \quad Q_*=\frac{Q}{M}\,,
\end{equation}
are constrained by the inequality 
\begin{equation}
a_*^2+Q_*^2\leq1\,.
\end{equation}
This is because the solutions with $1<a_*^2+Q_*^2$ do not describe 
black holes but exhibit, as a rule, one or more naked singularities.  
A  charged emitted particle with mass $m$ carries off $n$ units of the fundamental charge, say $\Delta Q=ne$,  and an angular momentum $-\Delta J=m$, both of order unity. Hence due to the 
constraint the number of charged particles needed to neutralize the 
hole is $Q/(n e)$, which is at most of order $M$, and the number of particles 
needed to spin down the hole is $J/m$ which is at most of order $M^2$. Thus 
the hole can discharge quickly~\cite{zaum74-247-530,gibb75-44-245,cart74-33-558} but the loss 
of angular momentum requires the same number of particles as the loss of mass. The 
question of the evolution of a rotating black hole was analyzed by 
Page in Ref.~\cite{page76-14-3260,page76-13-198} in great detail. By considering all the 
known particles with masses less than $20$ Mev, the temperature of a 
black hole with mass of order $10^{16}$ grams, he found that the emission 
of angular momentum increases greatly with $a_*$. Moreover, more than 
one half of the energy is emitted after $a_*$ reaches a small value of 
the order of $0.06$. From this point the power is within $1\%$ its 
Schwarzschild value and therefore the earlier assumption that decaying black 
holes have negligible rotation is valid. \\
These properties are challenged by the black holes of modified gravity, although the main argument should retain his strength, because the standard model Lagrangian, describing the matter part of the system, is not modified in the present considerations and the only additional particle in the gravitational sector (other than the massless graviton) is a massive scalar. In Page's times one did not consider the emission of dark matter particles, and we too avoid this question here.\\
But it is clear that the black hole will emit several species of massless and massive 
particles depending on his temperature. In this case the total luminosity of the black hole can be  
computed by summing over all particle species. 
Don Page was able to estimate the total power emitted: taking into account four kinds of neutrinos ($\nu_{e}$, $\nu_{\mu}$ and the two anti-neutrinos), 
the photon and the graviton, for $10^{17}g <M$ his result was 
\begin{equation}
P=2.28\times10^{-54}L_{\odot}(M_{\odot}/M)^2\,,
\end{equation}
where $81.4\%$ is in the four kinds of neutrinos, $16.7\%$ is in 
photons and $1.9\%$ is in gravitons. Here, 
$M_{\odot}=1.99\times10^{33} g$ is the solar mass and 
$L_{\odot}=3.9\times10^{33}$erg sec$^{-1}$ is the solar luminosity.
For $5\times10^{14}<M<10^{17}$ the black hole emits ultrarelativistic 
$e^{\pm}$ which may be treated as massless fermions, and the power is
\begin{equation}
P=4.07\times10^{-54}L_{\odot}
(M_{\odot}/M)^2\,,
\end{equation}
of which $45\%$ is in electrons and positrons, $45\%$ is in neutrinos, 
$9\%$ is in photons and only $1\%$ in gravitons. In all, most of the energy is radiated in the form of massless or nearly massless particles, as was o be expected on general 
grounds for a low temperature object. Moreover, the bulk of the radiation appears in standard model particles, rather than gravitons.

The black holes of modified gravity we are considering presently have a finite temperature and entropy and obey the 
first law of thermodynamics for a suitable defined thermodynamics energy. Thus they will radiate away their energy via Hawking steady emission for most of their lifetimes. 
They do this by emitting particles of the standard model, which is left untouched in modified gravity. Moreover, no new gravitational excitation are introduced except for a massive scalar,
so we may still consider the evaporation rate of black holes in modified gravity, for not too small masses, as substantially identical as for asymptotically flat black holes
in General Relativity\footnote{At least if we treat  gravity (modified or not) as a classical external field.}. 
Of course to fill in the details (like the precise percentages for example) deserves a more careful study. 

The conclusions we can draw from the above discussion and the examples we gave for the explicit solutions is that the simple scaling of the action grow with the thermodynamics energy 
is a direct consequence of the universality of Hawking radiation for a given particle spectrum.  In particular, it is largely independent on the gravitational sector and massive
states are not radiated anyway. So one expect the same complexity as in GR, if gravity is treated classically. 

One can see the connection with Hawking radiation quantitatively. In GR the Hamiltonian on a three-surface $\Sigma$ bounded by a sphere which is part of a horizon has the form
\begin{equation}
{\cal H}=\mathrm{bulk\;\,term}-\frac{1}{8\pi G_N}\int_{\cal M}(\kappa-
16\pi h^{-1/2}P^{ij}N_i\xi_j)dA +\mathrm{terms\;\,at\;\,infinity}\,,
\end{equation}
where $\kappa$ is the surface gravity, $N_{i}$ the shift, $\xi_{j}$ the normal to ${\cal M}$ within the three-surface
and we have reintroduced the Planck constant $h$. On shell the bulk term vanishes because it is a constraint, the momentum term also vanishes in a static geometry or when the shift is taken to vanish on the horizon, while the term at infinity is absent if $\Sigma$ is internal to the horizon.
Identifying the temperature $T=\kappa/2\pi$ and the entropy $S_{BH}=A/4G_N$ as usual, the action grow bound is $\dot{I}=TS_{BH}\leq 2E$, or by taking
derivatives\footnote{The time derivative of an inequality 
does not necessarily respect the inequality, but if the entropy \\increases with energy that is the case. The only exception would be systems with negative temperature.}
\begin{equation}
\dot{S}_{BH}\leq -2P/T\,,
\end{equation}
where $P$ is the power emitted by the BH. We have to consider now the entropy carried away by the Hawking radiation,  say $\dot{S}_{rad}$. By the generalized second law, 
\begin{equation}
0\leq \dot{S}_{BH}+\dot{S}_{rad}\,,
\end{equation}
so finally 
\begin{equation}
2P/T\leq \dot{S}_{rad}\,.
\end{equation}
We should note that one usually expects $P/T\leq \dot S_{rad}$ by conventional thermodynamics. 

If one accepts  Pendry's \cite{pendry-1983} universal bound on the entropy flow out of a thermal source radiating in vacuum (like the black hole)
\begin{equation}
\dot{S}_{rad}\leq \left(\frac{\pi P}{3}\right)^{1/2}\,,\label{PI}
\end{equation}
(we remember that $k_{B}=1$ in our units) then one gets for the power the limit,
\begin{equation}
\label{PB}
P\leq \frac{\pi T^2}{12}\,.
\end{equation}
As a qualification, the Pendry 
inequality holds only for outward flow of energy and therefore it does not represent the maximum rate of cooling of the black hole. This is easily
disposed off: since we took $P=-\dot{E}$, the left hand side of Eq.~(\ref{PB}) should be replaced by $\sum_{s}\bar{\Gamma}_{s}P_{s}$, where $\bar{\Gamma}_{s}$ the average over energy of the transmission coefficient of the potential barrier surrounding the black hole for a particle species $s$. It is a number of order one.   This is because the fraction $1-\sum_{s}\bar{\Gamma}_{s}P_{s}$ of the power is reflected back into the hole.
The left hand side is just the total power radiated via Hawking radiation by a Schwarzschild black hole, which saturates the inequality, and was used by Bekenstein~\cite{beken2001} 
long ago to infer the one-dimensional character of a black hole considered as an information transmission channel. In fact, the inequality (\ref{PI}) can be easily violated by 
transmitting over many parallel channels. The result for the action grow in the modified gravity models considered here indicates the validity of the same bound for the power emitted,
provided the power in these equations measures the rate of emission of the thermodynamical energy as defined in the text.\\
For different gravitational actions the particle spectrum sometimes changes radically. For example, adding a term $\sim R_{\mu\nu}R^{\mu\nu}$ will introduce a
massive spin two 
ghost, but what is the complexity of a negative norm state? However in one case the Weyl correction discussed above was in agreement with the standard complexity bound ($0<\epsilon\sigma$), in another it was not ($\epsilon\sigma<0$). It would be interesting to test whether in this case the theory contains ghost like excitations, which would violate the action grow bound.

\section{Conclusions}

In this paper we have investigated the general form of the action growth for some modified gravity
 black hole solutions. Within this more general framework different to the one of GR,
 new vacuum black hole solutions with non vanishing curvature may be found. In our analysis, we have considered several BH solutions where only one integration constant is present. Thus, 
 by making use of the First law of BH thermodynamics in these modified gravity models, we have shown  that the energy of our black holes is 
 always proportional to the integration constant associated with the solution.
 We should note that within the class of modified gravity models we are interested in, the First
 Law  can be derived from the equations of motion and making use of the  Killing
temperature and the Wald entropy and this fact seems to be a robust argument to substantiate the definition of the BH energy we have made use of.  In the case of solutions with constant Ricci curvature, 
we have confirmed the results of Refs.~\cite{Caibound, FRbound}, namely the action growth corresponds to 
the double of the Killing energy, in agreement with the result of Brown {\it et al.} in General Relativity~\cite{Susskind}. On the other 
hand, for solutions with non-constant Ricci curvature, the Kodama-Hayward BH energy emerges in the action growth. We recall  that the Kodama-Hayward energy is different to the Killing one due to the different expression of the
Kodama and Killing vectors associated with ``dirty'' BHs, and they coincide only when $g_{00}(r)g_{11}(r)=-1$. Our result is not surprising since the Hayward formalism is covariant
and valid for spherically symmetric dynamical space-times .

In the last part of our work, we considered a modified gravity model based on a Weyl-correction of gravity with an exact BH solution and we have derived the form of the related
action growth, which is still proportional to the Kodama-Hayward energy of the black hole itself. In one case the GR bound was satisfied, in another it was not. We argue
that the theory could contain a ghost like excitation. 

To interpret physically the obtained result, we argued on the basis of some black hole phenomenology that  the complexity bound as expressed by
the action grow is tightly related to the Hawking radiation process. Since the particle spectrum of $F(R)$-gravity is just the same as for GR, apart from a massive scalar, and the 
standard matter Lagrangian describing 
the matter sector is left untouched, the black holes of modified gravity radiate aways their mass in essentially the same way as in GR. In fact, writing the action in the Einstein frame, a scalar degree of freedom appears which is really a masked metrical invariant of the Jordan frame. This formulation of the theory has been studied elsewhere in the cited references, but for constant scalar field. Our description can be interpreted in the Einstein frame as the presence of a non constant scalar field. 

If the mass is defined as described in the text 
to represent the thermodynamical energy of the black holes, then the action grow must scale with this energy, as was actually found, and thus be the same as in GR up to numerical 
coefficients. This physical interpretation is not precise, since no detailed calculations were provided to fill in the details of the radiation process for general $f(R)$ models, except for the Brans-Dicke theory which has exactly the same black holes as General Relativity, as was shown by Hawking long ago.

\section*{Appendix A}
In this Appendix, we review some elementary aspects of  induced geometry associated with a $r$-constant surface. 
Let us start  recalling  the 4-dimensional metric,
\begin{equation}
 ds^2=g_{\mu \nu}dx^\mu dx^\nu=-\text{e}^{2\alpha(r)}B(r)dt^2+\frac{dr^2}{B(r)}+r^2s_{ab}(x^a)dx^adx^b\,,\label{AppA}
\end{equation}
where $s_{ab}(x^a)$ is a two dimensional ``horizon metric''.
Let denote by $n^\mu$ the unit normal vector to the surface $r$-constant, which reads
\begin{equation}
  n^\mu=(0, \frac{1}{\sqrt{g_{rr}}}, 0,0)=(0,\sqrt{B(r)},0,0)\,.\label{n}
\end{equation}
The induced metric $h_{\alpha \beta}(x^i)$ of a surface with constant $r$ is given by
\begin{equation}
h_{\alpha \beta}(x^i)=g_{\alpha \beta}(x^{\mu})-n_\alpha n_\beta\,,  
\end{equation}
namely
\begin{equation}
 dh^2=-\text{e}^{2\alpha(r)}B(r)dt^2+r^2s_{ab}dx^adx^b\,,
\end{equation}
and this may represent a time-like, space-like or null-like surface.
One has $\sqrt{-h}=r^2e^{\alpha}\sqrt{B}\sqrt{s} $, and the related  extrinsic curvature is defined as
\begin{equation}
K= \nabla_\alpha n^\alpha=h^{\alpha \beta}(x^i)\nabla_\beta n_{\alpha} \,.\label{K}
\end{equation}
Thus,  one obtains
\begin{equation}
K= \frac{\sqrt{B(r)}}{2} \left( \frac{1}{B(r)}\frac{d B(r)}{dr}+2 \frac{d \alpha(r)}{dr}  + \frac{4}{r}\right) \,,
\end{equation}
with scalar density 
\begin{equation}
\sqrt{-h}K= \sqrt{s}\frac{r^2\text{e}^{\alpha(r)} }{2} \left( \frac{d B(r)}{dr}+2 \frac{d \alpha(r)}{dr}  B(r)+ \frac{4 B(r)}{r}\right) \,.\label{AppAf}
\end{equation}
In our work, $\sqrt{s}=V_k$. A direct computation of the boundary term in (\ref{genBT}) leads to
\begin{equation}
BT=-2 \int_{\partial \mathcal M} d^3 x\sqrt{-h}F'(R)K= -V_k \Delta t \left[
F'(R)\text{e}^{\alpha(r)}r^2\left(\frac{d B(r)}{d r}+2 B(r)\frac{d\alpha(r)}{d r}+\frac{4 B(r)}{r}\right)
\right]\,,
\end{equation}
and we recover Eq.~(\ref{BT}).

\section*{Appendix B}

In this Appendix, we explicitly show that the equations of motion (\ref{one})--(\ref{two}) obtained by inserting the metric Ansatz (\ref{metric}) in the gravitational action
of $F(R)$-gravity are equivalent to the $(0,0)$- and $(1,1)$-components of the general field equations (\ref{keyeq}) of the theory (in the vacuum case, the other non-zero components, 
namely the
$(2,2)$- 
and $(3,3)$-components, are derived from the first two).

Let us rewrite Eq.~(\ref{keyeq}) as 
\begin{equation}
 F'(R)\left(R_{\mu \nu}-\frac{1}{2}R g_{\mu\nu}\right)+
\frac{1}{2}g_{\mu\nu}\left(RF'(R)
-F(R)
\right)
 -\left(\nabla_\mu \nabla_\nu-g_{\mu\nu}\nabla_\alpha \nabla^\alpha\right)F'(R)=0\,.
\end{equation}
The $(0,0)$- and $(1,1)$-components of this equation with the metric (\ref{metric}) read,
\begin{eqnarray}
 &&-\left(\frac{B(r)\text{e}^{2\alpha(r)}}{2r^2}\right)\left[r^2\left(RF'(R)-F(R)\right)-2F'(R)\left(k-B(r)-r\frac{d B(r)}{d r}\right)\right.\nonumber\\
& & \left.+2B(r)F''(R)r^2\left[\frac{d^2 R}{d
r^2}+\left(\frac{2}{r}+\frac{dB(r)/dr}{2 B(r)}\right)\frac{d R }{d
r}+\frac{F'''(R)}{F''(R)}\left(\frac{d R}{d
r}\right)^2\right]\right]=0\,,\label{unouno}
\end{eqnarray}
\begin{eqnarray}
 &&\left(\frac{1}{2B(r)r^2}\right)\left[
 r^2\left(RF'(R)-F(R)\right)-2F'(R)\left(k-B(r)-r\frac{d B(r)}{d r}\right)
\right. \nonumber\\&&
\left.
+4 F'(R)r B(r)\frac{d\alpha(r)}{dr}+
 F''(R)\frac{d R}{d r}\left(2B(r)r^2\frac{d\alpha(r)}{d r}+4B(r)r\right)
 \right]=0\,.\label{duedue}
\end{eqnarray}
Thus, Eq.~(\ref{unouno}) is equivalent to Eq.~(\ref{one}), while in order to obtain Eq.~(\ref{two}) we must substitute Eq.~(\ref{unouno}) in Eq.~(\ref{duedue}).

\section*{Appendix C}

In this Appendix, following Ref.~\cite{deruelle}, we will compute the boundary term for the Weyl model in (\ref{d action}) in the case of a SSS space-time.
First of all, we recall the general form of the boundary term for such a kind of theory in the form of a surface integral with $r$-constant, namely
\begin{equation}
BT=-2\int_{\partial \mathcal M} d^3 x \sqrt{-h}\Psi K\,, \label{BTgen}
\end{equation}
where we are using the parameterizations 
and the definitions in (\ref{AppA})--(\ref{AppAf}). Moreover, $\Psi$ is the trace of the tensor $\Psi^{ij}$,
\begin{equation}
\Psi^{ij}=-2h^{i\kappa}h^{jl}n^\mu n^\nu \phi_{\kappa\mu l\nu}\,,\quad \phi^{\kappa\mu l\nu}=\frac{d\mathcal L}{d R_{\kappa\nu l\nu}}\,,\label{psiij}
\end{equation}
where in our case
\begin{equation}
\mathcal L=\frac{1}{2\kappa^2}\left(R-2\Lambda+\sqrt{3}\sigma \sqrt{W}\right)\,. 
\end{equation}
One has
\begin{eqnarray}
\frac{\delta \mathcal L}{\delta R_{\mu\nu\xi\sigma}}&=& \frac{1}{2\kappa^2}\left\lbrace \frac{1}{2} (g^{\mu\xi}g^{\nu\sigma}- g^{\mu\sigma}g^{\nu\xi}) 
+ \frac{\sqrt{3} \sigma}{2\sqrt{W}} \,\times \right.\nonumber \\ \nonumber\\
& &\hspace{-20mm} \left.\left[ 2 R^{\mu\nu\xi\sigma} - (g^{\mu\xi}R^{\nu\sigma}+ g^{\nu
\sigma} R^{\mu\xi}- g^{\mu\sigma}R^{\nu\xi}- g^{\nu\xi}R^{\mu\sigma}) + \frac{1}{3}
(g^{\mu\xi}g^{\nu\sigma} - g^{\mu\sigma}g^{\nu\xi}) R \right]\right\rbrace \,.
\label{variation} 
\end{eqnarray}
By taking into account (\ref{n}) and the symmetries of the metric, it is easy to see that
\begin{equation}
\Psi\equiv h_{ij}\Psi^{ij}=-2 h_{00}h^{00}h^{00}n^r n^r h_{00}h_{11}h_{00}h_{11}\phi^{0101}=2\text{e}^{2\alpha(r)}\frac{d\mathcal L}{d R_{0101}}\,.
\end{equation}
A direct computation leads to
\begin{eqnarray} 
\left(\frac{\delta
\mathcal L}{\delta R_{0 1 0 1}}\right)&=& \frac{1}{4\kappa^2}
\left[ g^{00} g^{11} + \frac{\sqrt{3}
\sigma}{\sqrt{C^2}}\left(2 R^{0 1 0 1} - g^{00} R^{1 1} - g^{1
1} R^{0 0} + \frac{1}{3} g^{0 0} g^{1 1}
R\right)\right]\nonumber\\
&=&\frac{1}{4\kappa^2\text{e}^{2\alpha(r)}}(1-\epsilon\sigma)\,.
\end{eqnarray} 
We finally obtain,
\begin{equation}
 \Psi=\frac{1}{2\kappa^2}(1-\epsilon\sigma)\,,
\end{equation}
and from Eq.~(\ref{BTgen}) with (\ref{AppAf}) one has the result,
\begin{equation}
BT=-2 \int_{\partial \mathcal M} d^3 x\sqrt{-h}\left(\frac{1-\epsilon\sigma}{2\kappa^2}\right)K= 
 -\frac{V_k\Delta t}{2\kappa^2}(1-\epsilon\sigma)\text{e}^{\alpha(r)}r^2\left(
\frac{d B(r)}{d r}+2\frac{d\alpha(r)}{d r}B(r)+\frac{4 B(r)}{r}
\right)\,,
\end{equation}
which corresponds to (\ref{BTWeyl}).


\end{document}